\documentclass[aps,pre,twocolumn,showpacs]{revtex4}
\usepackage[T1]{fontenc}
\usepackage[latin1]{inputenc}
\usepackage{epsfig}

\makeatletter

\providecommand{\LyX}{L\kern-.1667em\lower.25em\hbox{Y}\kern-.125emX\@}

\makeatother

\begin{document}

\title{The interval
description of dynamics of celestial bodies in the planetary
problem}
\author{Valeriy~V.~Petrov }

\affiliation{Department of Radiophysics, Nizhny Novgorod
University, 23, Gagarin Avenue, 603950 Nizhny Novgorod, Russia }

\begin{abstract} The interval approach to computation of dynamics
of celestial bodies in the planetary problem has been considered.
It is based on the refusal from idealization of infinitely high
resolving capacity of measuring tools, and forms an absolutely
exact algorithm free of round-off error accumulation effect. The
possibilities of the proposed approach are shown by the examples
of Kepler's Problem and the problem of stability of the Solar
system major planets for time interval of 6 billion years. The
comparison of the interval and classical predictions of Kepler's
particle location in Kepler's orbit provides support for the
effect predicted by the theory, namely - conservation of the
interval within which the values of difference of interval and
classical coordinates lay with time. The computational results of
the Solar system major planet orbital dynamics agree with the
results obtained with the classical approach.
\end{abstract}


\maketitle

\section{Introduction}

The interval-discrete concept of dynamics of point systems has
been formulated in the paper \citep{b1}. The study is based on the
refusal of idealization of an infinitesimal error of observations
and calculations that is on the property of the limited resolving
capacity of measuring tools and measurement processing facilities,
which is not considered obviously in classical (trajectory)
concept of motion.\footnote{The idealization of an infinitesimal
error of measurements accepted in this concept is equivalent to an
assumption of existence of absolutely exact value of a physical
quantity. From the view of the information aspect it is equivalent
to the information infinity.} In the paper \citep{b2} the interval
description of macro bodies motion is applied to the planetary
problem. The numerical computations of the Solar system planetary
orbit evolution for time intervals of about 500-million years are
performed in the study with the help of interval equations. The
computations have verified the results obtained by classical
methods, and have enabled to make a conclusion that the proposed
interval approach is applicable for solving the problem of many
bodies.

This paper presents the continuation of the study having been
begun in \citep{b2}. The study objective is to completely realize
and show the advantages of the interval approach as applied to the
planetary problem and, thereby, to lay the groundwork for
experimental check of the equations of the interval theory.

In this connection we remind \citep{b1,b2} that formally the
interval equations are a system of integer mappings of
recurrent-type obtained by a special procedure of quantization of
the time and spatial continuums of the system and the intervals of
its dynamic variables. The feature of these mappings lies in the
fact that they form an absolutely exact computational algorithm
free of the effect of the round-off error accumulation. In this
case it is possible to carry out computations with an unlimited
number of iterations. The other feature of these mappings is a
reasonable simplicity significantly reducing amount of
computations in comparison with the classical approach. In the
paper the mentioned features are illustrated by the examples of
Kepler's Problem and the problem of stability of the Solar system
for cosmogonic times.

Kepler's Problem is of interest as a basic problem. In particular,
it allows demonstrating the interval tube effect \citep{b2}. Its
manifestation is that the values of deviations of variable
interval centers from their classical values lies in a strictly
fixed interval the width of which does not exceed the initially
preset width of interval variables. Classically it means that
precision of the interval prediction for Kepler's particles
dynamics remains a constant value irrespective of "integration"
duration. It is natural that the real reliability of such
prediction will be limited by the time of system's fall outside
the limits of its «horizon of predictability» \citep{b3}. In this
connection the interval dynamics in contrast to the classical
dynamics in which there is no concept of "horizon of
predictability» at all, allows to exactly estimate the "lifetime"
of theoretical prediction, which is always finite in practice.

Kepler's Problem allows demonstrating one more important property
of the interval dynamics. That is ability of the interval tube for
closing if the corresponding classical trajectory represents a
closed phase curve. The uniqueness of this property is manifested
by its inhesion to particularly numerical result, i.e. it means an
absolutely exact recurrence of numerical parameters of the tube
through one or several turns of a particle.

And, at last, there is one more interesting effect, which is
demonstrated by the example of Kepler's Problem. An interval
particle alongside with an orbital moment has its own kinetic
moment (spin). This effect is one of consequences of the interval
nature of physical system space. It is conditioned by the fact
that an interval particle in contrast to a classical one has the
status of not a mathematical point, but a physical one. In other
words the particle has finite size (determined by the width of
intervals of the problem spatial variables) and, as a consequence,
possesses its own moment of an impulse.

Freedom of the effect of round-off error accumulation allows
successful solving of particularly academic problems alongside
with practical ones. One of the problems is a study of a dynamic
stability of the Solar system major planets. Difficulty of its
solution is caused, first of all, by long-term character of the
required prediction of the planets motion assuming carrying out
the computations with time intervals of the Solar system age
order. In the present work such computations are made for a time
interval of six billion years. The obtained results agree with the
results obtained earlier within the limits of the classical
approach as for external planets \citep{b7,b6,b4,b5}, so as for
internal planets \citep{b7,b4,b5}.

\section{Kepler's problem }

Let's consider an interval statement of Kepler's Problem in the
space of polar coordinates presenting Hamiltonian of a classical
analogue of the system under consideration in the following form:
\begin{equation}
    H=\frac{1}{2}(p_{r}^{2}+\frac{p^{2}_{\varphi}}{r^{2}})-\frac{\mu}{r}.\\
    (p_{\varphi}=c=\textrm{const})
\end{equation}
In order to take into account the limited resolving capacity of
the instrumental observation facilities and to pass to the
interval description of the system dynamics (1), we introduce
following to \citep{b1} quantized spaces of its variables in the
form of lattices with the following periods: $\rho$ - for radius
$r$, $\sigma$ - for angle $\varphi$, $\upsilon$ - for impulse
$p_{r}$ and $\vartheta$ - for time $t$. The designated periods
will characterize the resolving capacity of corresponding
observation procedures at the theoretical level. Thus the motion
of a particle is described by not real numbers, but integer
interval of coordinates $R_{n}=R(T_{n}), \Phi_{n}=\Phi(T_{n})$,
impulse $P^{r}_{n}=P_{r}(T_{n})$ and time $T_{n}$ in the following
form
\begin{equation}
\begin{array}{l}
    R_{n}=r_{n}+[-N,N],\\
    \Phi_{n}=\varphi_{n}+[-N,N],\\
    P^{r}_{n}=p_{n}+[-N,N],\\
    T_{n}=[nN,nN+2N],\\
\end{array}
\end{equation}
where $n=0,1,2,...$ - the number of the temporary variable lattice
site with period $\tau=\vartheta N$, and $N\gg1$ - the interval
(integer) number.

According to (2) the state of the considered system is localized
not in a point but in a multidimensional interval with the
following absolute values of the half-width: $\alpha_{r}=\rho N$ -
on variable $r$, $\alpha_{\varphi}=\sigma N$ - on variable
$\varphi$, $\beta_{r}=\upsilon N$ - on variable $p_{r}$ and $\tau$
- on variable $t$. In terms of classical representations these
values of half-width can be interpreted as the characterization of
the numerical description of system. At that the given "precision"
does not vary while the specified multidimensional interval moves
with time in the quantized phase space forming an interval tube.
Later we shall show that every such tube contains at least one
classical trajectory.

As follows from the theory  \citep{b1}, the designated values of
the half-width are connected with the maximum velocities of change
of the system generalized coordinates and impulses by ratios
\begin{equation}
    \alpha_{r}=\tau V_{r},\\ \alpha_{\varphi}=\tau V_{\varphi},\\ \beta_{r}=\tau W_{r},
\end{equation}
where $V_{r}\geq|dr/dt|_{\textrm{max}},\
V_{\varphi}\geq|d\varphi/dt|_{\textrm{max}},\
W_{r}\geq|dp_{r}/dt|_{\textrm{max}}$. At that the equations of
dynamics for the integer centers $r_{n},\varphi_{n}$ and $p_{n}$
of the intervals (2) can be written down in the form of \citep{b9}
\begin{equation}
\begin{array}{l}
    r_{n+1}=r_{n}+[w\tilde{p}_{n}],\\
    p_{n+1}=[\tilde{p}_{n+1}],\\
    \varphi_{n+1}=[\tilde{\varphi}_{n+1}],
\end{array}
\end{equation}
where
\[
\tilde{p}_{n+1}=p_{n}+\frac{\tau}{\upsilon}
\left(-\frac{\mu}{\rho^{2}r^{2}_{n}}+
\frac{c^{2}}{\rho^{3}r^{3}_{n}}\right),
\]
\[
\tilde{\varphi}_{n+1}=\varphi_{n}+\frac{\tau}{\sigma}
\left(\frac{c}{\rho^{2}r^{2}_{n}-I}\right),
\]
\[
I=I_{0}+\delta \sum_{m=0}^{n}\sin(\sigma\varphi_{m}),
\]
\[
w=W_{r}\tau/V_{r}.
\]
Here symbol $[x]$ \citep{b1,b2} means the procedure of a round-off
of real number $x$ up to the nearest integer, and function $I$
where  $I_{0}$ and $\delta$ are the constants defined by the
problem intervalization parameters, means the inertia moment of an
interval particle. The occurrence of $I$ in the equation for
$\varphi_{n}$ of the system (4) is caused by the fact that the
interval particles have spin, that, in its turn, and as is
mentioned above, is a consequence of finiteness of the
localization area size of the interval particles. The detailed
discussion of this effect is beyond the problem of this work. It
will be an objective of a special paper.

Let's apply the equation system (4) to calculate the motion of the
particle over the closed orbit taking for definiteness $\mu=1,
c=1, e=0.3$ ($e$ - eccentricity). Perform the calculations with
step $\tau=0.07238$ that is $\sim$ 0.01 of the motion period over
the selected orbit. Take $N=10^{7}, V_{r}=0.303, W_{r}=0.52,
V_{\varphi}=1.71, I_{0}=1.3\times10^{-4}, \delta=1.5\times10^{-5}$
for the specified parameters. In this connection according to (3)
the half-width of the interval of particle localization in the
space shall not exceed $\alpha_{r}\approx0.022$.

The results of calculations are presented in Figures 1-3. Fig. ~1
shows the performance of full energy conservation laws and the
kinetic moment of system. The figure presents the relative
fluctuations of current energy values $\Delta E/|E|$ and kinetic
moment $\Delta c/c$ for time interval of $2.4\times10^{6}$
periods. These fluctuations are computed with the following
formulas:
\[
\Delta E=E_{\textrm{int}}-E,
\]
\[
\Delta c=c_{\textrm{int}}-c,
\]
where
\[
E_{\textrm{int}}=\frac{1}{2}\left(\upsilon^{2}p^{2}_{n}+
\frac{c^{2}}{\rho^{2}r^{2}_{n}}\right)-\frac{\mu}{\rho r_{n}},
\]
\[
c_{\textrm{int}}=\frac{\rho^{2}r^{2}_{n}}{\rho^{2}r^{2}_{n}-I}c, \\\
E=-\frac{\mu^{2}(1-e^{2})}{2c^{2}}.
\]
It follows from Fig. 1 that the designated fluctuations lay in
strictly fixed intervals which do not vary with the course of
time, and equal by the width to $\sim 0.008$ - for energy, and to
$\sim 10^{-4}$ - for kinetic moment.

The Fig.~2 shows effect of an interval tube. Fig. 2 presents the
time dependence of $\Delta x=x_{\textrm{int}}-x_{\textrm{kep}}$
and $\Delta y=y_{\textrm{int}}-y_{\textrm{kep}}$ characterizing
the deviations of coordinates of the interval tube axis
$x_{\textrm{int}}=\rho r_{n}\cos(\sigma\varphi_{n})$,
$y_{\textrm{int}}=\rho r_{n}\sin(\sigma\varphi_{n})$ on the
corresponding values of Kepler's coordinates
\[
x_{\textrm{kep}}=\frac{c^{2}}{\mu}\frac{\cos(\sigma\varphi_{n})}{1+e\cos(\sigma\varphi_{n})},
\]
\[
y_{\textrm{kep}}=\frac{c^{2}}{\mu}\frac{\sin(\sigma\varphi_{n})}{1+e\cos(\sigma\varphi_{n})}.
\]
According to Fig. 2 these deviations do not fall outside the
limits of the fixed intervals with the half-width
$\alpha_{x}\approx0.006$ and $\alpha_{y}\approx0.012$, i.e. the
specified values satisfy the condition of localization
$\alpha_{x,y}<\alpha_{r}$. This implies that the classical
(Kepler's) trajectory all over does not fall outside the limits of
the intervals predicted by the equations (4), and entirely lies
inside the corresponding interval tube.

One more interesting property of the interval tube becoming
apparent in case of periodic motions is its closability. In the
example under analysis such closing is observed on a phase plane
$r_{n}, p_{n}$. As calculations show, for initial conditions
\begin{equation}
    r_{0}=350790178,\ p_{0}=-20
\end{equation}
the closing of the interval tube $R_{n}, P^{r}_{n}$ (interval
analogue of the classical trajectory $r, p_{r}$) occurs in 100
steps, i.e. in one turn of a particle. This result is illustrated
in Table 1 where the values of quantum numbers $r_{n}$ and $p_{n}$
are calculated by means of (4) in the beginning of the motion and
in 100 steps.

\begin{figure}
\centerline{\epsfig{file=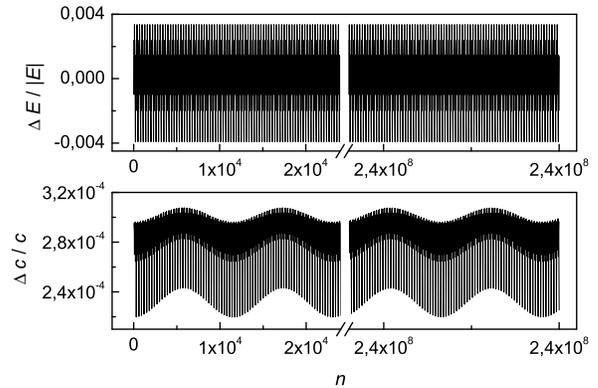, width=1.0\columnwidth}}
 \caption{The time dependence between the relative quantities of full energy fluctuations
 ($\Delta E/|E|$) and the orbital kinetic moment ($\Delta c/c$) of Kepler's particle.}
  \label{fig1}
\end{figure}

\begin{figure}
\centerline{\epsfig{file=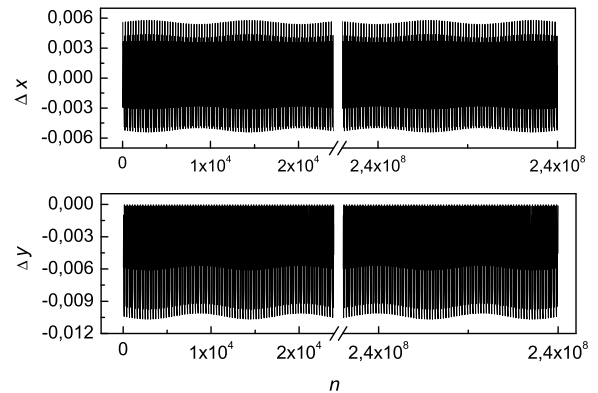, width=1.0\columnwidth}}
 \caption{The time dependence between the differences of the interval
 and classical values of Cartesian coordinates of Kepler's particle.}
  \label{fig2}
\end{figure}

\begin{table}
\centering
 \begin{minipage}{50mm}
  \caption{The time dependence of phase variables $r_{n}$ and $p_{n}$ on a periodic trajectory.}
  \begin{tabular}{@{}llrrrrlrlr@{}}
  \hline
   $n$ & $r_{n}$ & $p_{n}$ \\
 \hline
 0 & 350790178 & -20 \\
 1 & 352000368 & 9742590 \\
 2 & 354394531 & 19274116 \\
 ..... & ................ & .............. \\
 100 & 350790178 & -20 \\
 101 & 352000368 & 9742590\\
 102 & 354394531 & 19274116\\
 \hline
\end{tabular}
\end{minipage}
\end{table}

As is obvious, coincidence of these values represents not an
approximate result, but an absolutely exact numerical result. Such
property is a distinctive feature of the motion interval
description and in principle is not realized within the limits of
classical calculation means.

The picture of the interval description of particle motion in
Kepler's Problem is not complete if not to concern one more aspect
of the interval theory, namely, an estimation of particle position
predictability horizon. As follows from \citep{b1}, occurrence of
such characteristic as "lifetime" of the theoretical prediction in
the interval theory is a direct consequence of the explicit
accounting of the interval nature of physical system phase space.
The elementary cell of such space contains not one but $2^{2s}$
points in the form of combinations of integer values of impulses
and coordinates, the single-type variables in which can differ
from each other by not more than one quantum. The beam $2^{2s}$ of
interval tubes coming out of such cell (physical point) gives the
reliable prediction of the current system state till the
integration of intervals of all phase variables for each of
variables keeps in the interval of width $4N$. At violation of
this condition if only for one of the variables it is possible to
speak about an output beyond the horizon of the system state
predictability.

This is the formality of the problem. In order to characterize the
problem practically, let us consider particular procedures carried
out when estimating the horizon of Kepler's particles position
predictability. And by the example of these procedures we shall
demonstrate technology.

Let's previously notice that the phase space of the considered
system represents section $p_{\varphi}=c$ and forms a
three-dimensional lattice with periods $\rho, \upsilon$ and
$\sigma$ for phase variables $r_{n}, p_{n}$  and  $\varphi_{n}$,
accordingly. Hence, the elementary cell of such space is a cube
with a one-quantum side. Let's introduce an estimation of
divergence of a beam of the trajectories coming out of this cell
during the initial moment of time. We analyze all the possible
combinations of trajectory pairs, and select that very pair in
which the trajectories diverge with the maximum velocity. In this
case we calculate the divergence time comparing coordinates
$x_{\textrm{int,1}}, y_{\textrm{int,1}}$ and $x_{\textrm{int,2}},
y_{\textrm{int,2}}$ of each pair applying criterion
\begin{equation}
    \Delta l\leq l_{\textrm{loc}},
\end{equation}
where
\[
    \Delta l=\sqrt{(x_{\textrm{int,1}}-x_{\textrm{int,2}})^{2}+
    (y_{\textrm{int,1}}-y_{\textrm{int,2}})^{2}},
\]
and $l_{\textrm{loc}}\approx0.024$. This value of width of the
interval of the spatial localization of the particle is selected
on the basis of the computation results presented in Fig. 2.

As the first example consider the dynamics of $\Delta l$ for a
beam of trajectory tubes coming out of the cell, one top of which
corresponds to the initial condition (5). For this condition the
result of computation of the time dependence of value $\Delta l$
is shown in Fig. ~3{a} and corresponds to the pair
\begin{equation}
\begin{array}{l}
    r_{01}=350790178,\ p_{01}=-20,\ \varphi_{01}=-20,\\
    r_{02}=350790179,\ p_{02}=-19,\ \varphi_{02}=-20.
\end{array}
\end{equation}
It follows from Fig. 3{a}, because of (6), that the particle
position predictability horizon is in the order of
$2.5\times10^{5}$ periods of the orbit motion.

Estimating this result, it is well to bear in mind that it
corresponds to the localization of the initial particle state in
terms of a phase space elementary cell. In practice such high
degree of localization is not always achievable. Taking into
consideration the real possibilities of observation tools, we
should expand the area of the system initial state localization to
some set of elementary cells.\footnote{Note that in this case all
cells of the given set of cells are invested with the property of
empirical indistinguishability from one another, and that at the
level of formal-mathematical means is described applying the
tolerance relation \citep{b8,b9}.} Such roughening of the system
description leads as a rule to reduction of the dynamics reliable
prediction time. This effect is shown in Figs. 3{b} and 3c. At
that Fig. 3{b} describes pair
\begin{equation}
\begin{array}{l}
    r_{01}=350790168,\ p_{01}=-20,\ \varphi_{01}=-20,\\
    r_{02}=350790169,\ p_{02}=-19,\ \varphi_{02}=-20,
\end{array}
\end{equation}
and Fig. 3{c} describes pair
\begin{equation}
\begin{array}{l}
    r_{01}=350790228,\ p_{01}=-30,\ \varphi_{01}=-20,\\
    r_{02}=350790227,\ p_{02}=-29,\ \varphi_{02}=-20.
\end{array}
\end{equation}
The elementary cell of pair (8) is ten (10) quanta apart from the
cell of pair (7) over $r_{n}$. This cell is characterized by an
essentially smaller predictability horizon (if compare with (7))
equal to $\sim7\times10^{3}$ periods of the particle. We have
still smaller predictability horizon of $\sim4.5\times10^{3}$
periods for pair (9) being 50 quanta apart from pair (7) over
$r_{n}$ and 10 quanta apart - over $p_{n}$.

\begin{figure}
\centerline{\epsfig{file=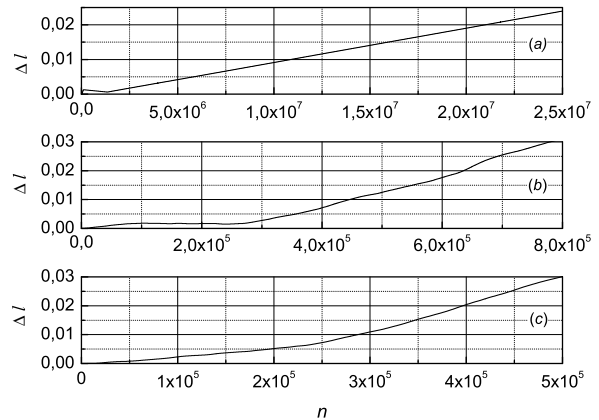, width=1.0\columnwidth}}
 \caption{The time dependence of $\Delta l$ distance between Kepler's particle locations belonging
 to two phase trajectories physically imperceptible in the initial moment.}
  \label{fig3}
\end{figure}

\section[]{ORBITAL DYNAMICS OF PLANETS OF SOLAR SYSTEM }

With the help of the interval equations \citep{b2} numerical
computation of motion of the major planets of the Solar system in
the time interval of $6\times10^{9}$ years has been made.
Computations have been made with step $\tau=0.012$ year for
$N=10^{9}$ . The masses of the planets have been taken from the
system of constants IAU 1964. The initial rectangular heliocentric
coordinates and velocities are related to the equator and
correspond to the stage of 1949, Dec. 30.0 ET=JED 2433280.5. To
study a long-term behavior of the planet orbits and their possible
drift in a chaos zone, the time dependence of maximum
eccentricities and inclinations \footnote{Inclinations of orbits
of all planets have been calculated in the heliocentric equatorial
system of coordinates.} of orbits for the interval centers of the
mentioned variables has been calculated. The values of the
half-widths of these intervals ($\alpha_{e}$ - for eccentricity
and $\alpha_{i}$ - for inclination) are presented in Table ~2. The
maximum values have been selected from the set of values in time
intervals of 6 million years, i.e. the technique similar to
\citep{b7} has been applied. The results of the calculations are
presented in Figures 4-7.

\begin{table}
\centering
 \begin{minipage}{50mm}
  \caption{The values of half-widths of intervals of planetary orbit
  eccentricity ($\alpha_{e}$) and inclination ($\alpha_{i}$).}
  \begin{tabular}{@{}llrrrrlrlr@{}}
  \hline
   $ $ & $\alpha_{e}$ & $\alpha_{i}$ \\
 \hline
 Mercury & 0.17 &  \\
 Venus & 6.5E-3 &  \\
 Earth & 0.0165 &  \\
 Mars & 0.025 & $\leq$5.0E-5 \\
 Jupiter & 0.035 & \\
 Saturn & 3.5E-3 & \\
 Uranus & 3.0E-3 & \\
 Neptune & 4.5E-3 & \\
 Pluto & 0.045 & \\
 \hline
\end{tabular}
\end{minipage}
\end{table}

Figs. 4 and 5 contain graphs of the above-mentioned dependences
for the internal planets. They show a possible drift of orbit
element values. Therewith the growth of the Mercury eccentricity
(for which the zone of chaos is maximum) is limited (according to
Fig. ~4 and Table ~2) by an interval of values the upper limit of
which does not exceed $\sim$0.38. For the considered time interval
this result agrees with the results obtained in \citep{b7,b4,b5}.

\begin{figure}
\centerline{\epsfig{file=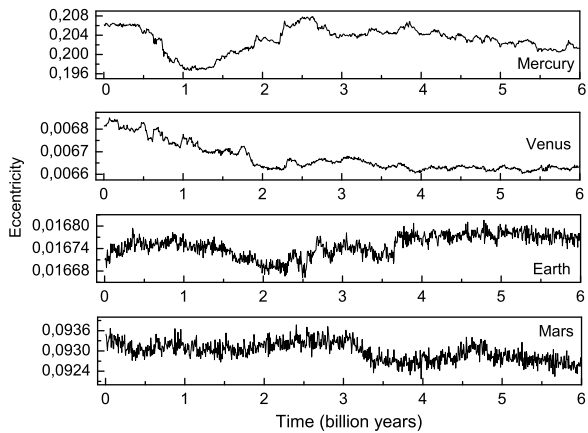, width=1.0\columnwidth}}
 \caption{The time dependence of maximum orbit eccentricities of internal planets (for interval centers).}
  \label{fig4}
\end{figure}

\begin{figure}
\centerline{\epsfig{file=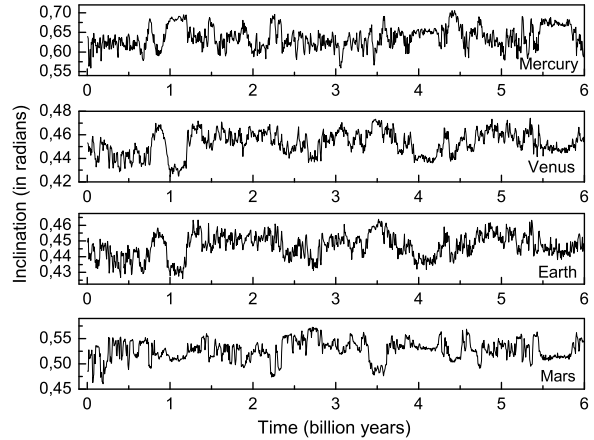, width=1.0\columnwidth}}
 \caption{The time dependence of maximum orbit inclinations of internal planets (for interval centers).}
  \label{fig5}
\end{figure}

\begin{figure}
\centerline{\epsfig{file=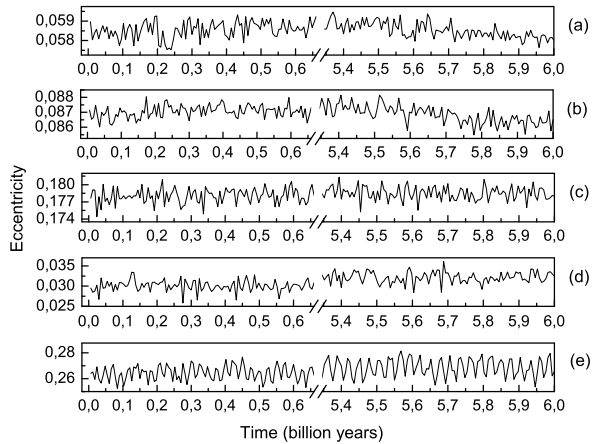, width=1.0\columnwidth}}
 \caption{The time dependence of maximum orbit eccentricities of external planets (for interval centers)
 : (a) - Jupiter, (b) - Saturn, (c) - Uranus, (d) - Neptune, (e) - Pluto.}
  \label{fig6}
\end{figure}

\begin{figure}
\centerline{\epsfig{file=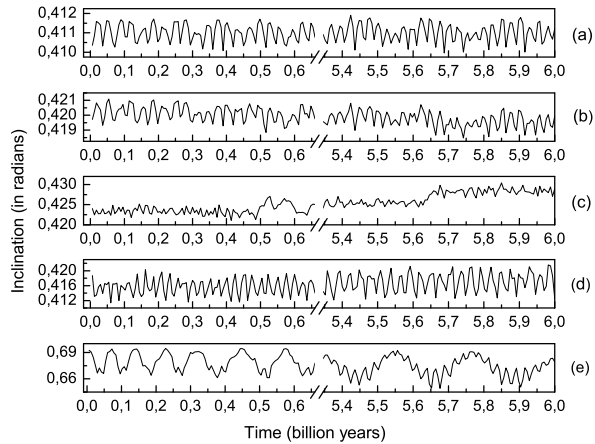, width=1.0\columnwidth}}
 \caption{The time dependence of maximum orbit inclinations of external planets (for interval centers)
 : (a) - Jupiter, (b) - Saturn, (c) - Uranus, (d) - Neptune, (e) - Pluto.}
  \label{fig7}
\end{figure}

Computation of behavior of the external planets (Figs. 6 and 7)
also coincides with that having been obtained earlier by the
authors having been mentioned and also by \citep{b6}. The Uranus
is an exception. For the Uranus, as seen from Fig. ~6{c}, value
$e_{max}$(with regard to the interval correction of Table 2) comes
up to value of $\sim$0.178. Thus $e_{min}\approx2\times10^{-2}$
and as computations show, the Uranus eccentricity drift in this
range has a random (stochastic) character, i.e. it can be regarded
as a feature of chaotization of its motions and a motion of the
external planets as a whole. A source of such chaotization is
overlapping of the components of triple resonance of average
motions of the Jupiter, Saturn and Uranus having been analyzed in
the work of \citep{b10}. Other source is overlapping of resonant
areas in the vicinity of the Uranus and Neptune orbits analyzed in
\citep{b11,b12}. At the same time, the obtained result is still
not sufficient for final conclusions. To get an unambiguous answer
about the nature of motion of the external planets, additional
computational investigations are necessary.

Completing the description of the results of this part of work,
let us touch upon the problem of labor coefficient of the
conducted computations. They have been carried out with the help
of a personal computer with processor AMD Athlon 64 3400+. To
improve the reliability, the computations have been conducted with
extended precision having demanded the doubling of computation
time. This time has been 1000 hours for time interval of 6 billion
years ($5\times10^{12}$ steps).

\section{Conclusions}

The conducted computations show the efficiency of the interval
approach, its adequacy to the problems of celestial bodies'
dynamics. The interval means of motion description assure to
obtain a solution in the defined strictly fixed interval of
divergence from the classical trajectory (in case of periodic
motions) and to attribute a unique property of absolutely exact
closing. In solving the problems of long-term motion prediction
the interval approach has one more advantage. Being free of
round-off error accumulation effect it enables to computationally
investigate the dynamics of planets in an arbitrary large time
interval. At that the interval theory includes special
computational procedure for estimating the horizon of the motion
investigated aspects predictability.

The examples considered in the paper do not obviously solve all
the problems related to the interval planetary dynamics. Along
with academic problems, a set of problems being solved thanks to
the creation of highly-precise ephemerides of planets and
satellites is particularly actual.  Moreover, the potential of the
interval approach can be realized with the greatest efficiency in
the class of applied problems.

It is obvious, that such realization is impossible without
interest on the part of experts belonging to the appropriate
application areas. And one of the problems of the present
publication consists in turning the experts' attention to
perspectiveness of application of the interval theory methods and
algorithms.

\label{lastpage}

\end{document}